# Room-temperature strong coupling at the nanoscale achieved by inverse design

**Yael Blechman[1], Shai Tsesses[1], Matthew Feinstein[2,3], Guy Bartal[1] and Euclides Almeida[2,3]**

[1]*Andrew and Erna Viterbi Department of Electrical Engineering, Technion – Israel Institute of Technology, Haifa 32000, Israel*
[2] *Department of Physics, Queens College, City University of New York, Flushing, New York 11367, USA*
[3] *The Graduate Center of the City University of New York, New York, New York 10016, USA*
Author e-mail address: yaelble@campus.technion.ac.il

**Abstract:** Room-temperature strong coupling between plasmonic nanocavities and monolayer semiconductors is a prominent path towards efficient, integrated light-matter interactions. However, designing such systems is challenging due to the nontrivial dependence of the strong coupling on various properties of the cavity and emitter, as well as the subwavelength scale of the interaction. In this work, we develop a methodology for obtaining hybrid nanostructures consisting of plasmonic metasurfaces coupled to atomically thin $WS_2$ layers, exhibiting extreme values of Rabi splitting, by inverse design of the near-field plasmonic response. Contrary to common measures such as the quality factor or the mode volume, our method relies on an overlap-integral-based metric. We experimentally measure large values of Rabi splitting for our nanoantenna designs, while providing theoretically optimal configurations for several additional types of nanostructures. Our results open a path to maximizing light-matter interactions in integrated platforms, for classical and quantum-optical applications.

## Introduction

The field of nano-optoelectronics aims at finding new possibilities for light-matter interaction and control, utilizing research advances in nanophotonics-related design and fabrication capabilities. A prominent manifestation of such interaction is the strong coupling (SC) between an optical resonator and a radiation source (emitter) which results in creation of hybrid modes that have simultaneous light-like and matter-like properties. While SC was originally observed in gases and at cold temperatures, recent fabrication methods have enabled its realization in solid state systems at room temperature. In particular, the hybrid structures consisting of excitonic emitters in certain two-dimensional materials and plasmonic metasurface (MS) cavities can provide robust room-temperature SC in nanoscale open cavities, opening an active area of research[1–4].

Transition-metal dichalcogenides (TMDs) are 2D direct-bandgap semiconductors with large oscillator strength and binding energy, turning these excitonic emitters into promising candidates for SC. Their transition energy is in the Visible-IR range, which, combined with their high mobility and robust and fast response, makes them adequate for a variety of light-matter interactions, and an excellent choice for development of photodetectors, modulators and other opto-electronic devices[5–7].

On the other hand, plasmonic metamaterials are well-established building blocks for control and enhancement of light in the nanoscale[8,9]. These flat, open cavities are only a few tens of nm thick and exhibit high field confinement beyond the diffraction limit, resulting in pronounced local field enhancement. In addition to the field localization, the optical response of plasmonic MS is versatile and can be controlled on sub-wavelength scales. The hybridization of plasmonic metasurface and TMD structures can therefore significantly impact integrated nanophotonics in terms of ease of design, fabrication and- above all - improved performance of devices[10,11].

Indeed, SC at room temperature in plasmonic-TMD hybrid structures was experimentally demonstrated by several groups[2,6,12–17], relying on the "spectroscopic approach", where the plasmonic resonance frequency of the MS is matched with the excitonic transition of the TMD by tuning design parameters (usually a single one). However, such an approach is not always sufficient to produce an optimal value of SC.

Here, we achieve state of the art values for SC in hybrid 2D-plasmonic structures, by developing a systematic, general and extensible inverse design technique which utilizes the plasmonic control to fully harness the potential of nanoscale SC via optimization of their near-field response. Our method can be applied to various types of MS such as nanohole arrays, antenna arrays or single nano-particles, providing optimal SC for each particular setup. We validate our nano-antenna design with measurements. Furthermore, our optimization function can be combined with additional plasmonic phenomena derived from the near-field response, such as multi-resonant behavior, directionality, incoming polarization selectivity, focusing, plasmonic laser, enhanced nonlinearity and more, leading to new functionalities.

## Nanoscale near-field inverse design for SC

A typical inverse design process consists of defining a Figure of Merit (FOM) and using an optimization procedure to find the parameters to design a structure that maximizes the FOM[18].

Although exciton-photon interaction is essentially a quantum mechanical phenomenon, it can be described classically by a set of coupled harmonic oscillators[1,6,19,20] and, consequently, a numerical simulation based on Maxwell's equations can be used to accurately predict the Rabi splitting in a hybrid structure (Figure 1a). The new eigenmodes, the lower and upper polaritons (LP/UP), exhibit the typical anti-crossing behavior (Figure 1b). At the anticrossing point (when the resonant frequencies of the MS and the exciton are equal), the spectral distance between the eigenmodes equals the strength of the coupling (in units of energy). Therefore, the coupling strength cannot be, in general, deduced from a single far-field spectral response, rendering a direct optimization for SC impractical.

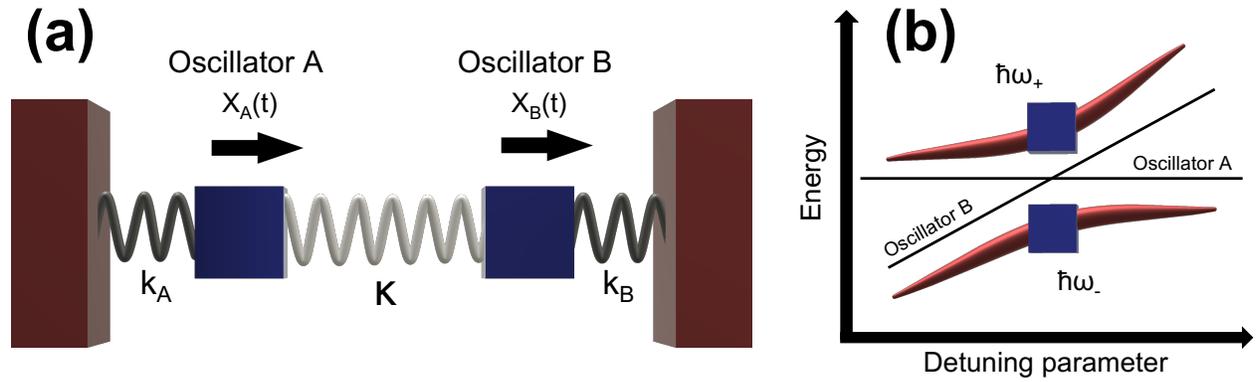

**Figure 1 (a)** The schematic representation of the strong coupling phenomenon, which can be described classically by a coupled oscillator model. **(b)** The resulting avoided crossing behavior of the hybridized mode frequencies $\hbar\omega_+, \hbar\omega_-$.

It is well known that large SC interaction requires small mode volumes and large Q factors[17,21]. However, it is not straightforward to define mode volume of a periodic open nanoscale plasmonic cavity, and even harder to implement in an optimization process for practical prediction of SC[22]. In addition, both the plasmonic field and the excitonic material are highly localized, but the mode volume does not take into account their overlap - which might vary between different setups, and therefore a single number is not sufficient to describe the cavity. This becomes especially important for nanometric dimensions, for example when the complicated spatial pattern of plasmonic hot-spots (confined in 3D) interacts with either an infinite 2D emitter material (confined in a single dimension) or a quantum dot (also 3D-confined).

## Results

Our main contribution in this paper is developing a FOM for achieving large SC. Our FOM is based on the near field of the plasmonic MS exactly at the **spatial location** of the excitonic 2D material, where the interaction takes place -- rather than on the far field spectral response of the MS. Indeed, the coupling strength from an excitonic emitter in a plasmonic cavity can be evaluated as[1]

$$\hbar\Omega \propto d \cdot E,$$

where $d$ is the dipole moment of the emitter, and $E$ is the electric field at the emitter's location in the excitonic frequency. Since in a monolayer 2D material the dipoles are mostly in plane [23,24], the scalar product $\boldsymbol{d} \cdot \boldsymbol{E}$ at each point $(x, y)$ in the monolayer plane is approximately equal to the product of the scalar quantities $d \cdot E_x$, where the incident polarization is assumed to be in the $x$ direction. To estimate the overall interaction strength, we integrate the contributions from each spatial location in the monolayer[25]. Additionally, in order to prioritize the creation of sharp, field enhanced features ("hot-spots"), we require that the minimal (normalized) field intensity used for the computation be larger than a certain threshold $t$. Increasing the threshold will generally lead to fewer very bright and small hot-spots, while decreasing the threshold will lead to eventual disappearance of the hot-spots. We find that in plasmonic-TMD structures, the value $t = 1.5$ provides a good tradeoff between these characteristics. The resulting FOM which should predict the magnitude of the SC in a particular setup is, therefore:

$$E_m(\omega) = \iint_{|E_x(x,y)|>t} |E_x(\omega;x,y)|dxdy, \quad E_m = \frac{1}{n}\sum_{i=1}^{n} E_m(\omega_i),$$

where the average is computed over a range of frequencies around the excitonic transition (to make the optimization more stable).

In our work, we use the Finite-Difference Time-Domain (FDTD) method for numerical simulation of the EM response, to calculate the value of $E_m$ for a given configuration, together with genetic algorithm optimization for finding the values of the parameters defining the MS geometry which maximize $E_m$ (see Supplementary Material). The genetic algorithm we use is efficient for objective functions which are not necessarily continuous or differentiable, in contrast to other methods such as steepest descent[18,26].

We implement the inverse design process on several different types of nanostructures, obtaining plasmonic MS exhibiting strong near-field enhancement and hot-spots, by optimizing the geometry with respect to the $E_m$ function.

For nanoantenna array geometry, the parameter space consists of antenna length, width, pitch and array periodicities (Figure 2a). The spatial location of the near-field computation is shown as dashed red line, representing the location of the WS$_2$ monolayer in the corresponding hybrid structure. The optimal MS configuration is resonant at the excitonic frequency (Figure 2b), while the near field distribution exhibits strong hot-spots (Figure 2c). The hybrid plasmonic-TMD structure possesses large SC response, extracted from the plots of the reflectance as a function of the detuning parameter (in this case, antenna short side), both simulated (Figure 2d,e) and measured (Figure 2f). The simulations of the hybrid structures were performed via 3D FDTD, fitting the dielectric function of WS$_2$ to experimental data[27].

For nanohole geometry, we consider arrays of rectangular and circular hole shapes milled in a gold layer with the corresponding parameter spaces (Figure 3a-c). The 3nm-thick WS$_2$ flake in the hybrid structures is placed directly on top of the MS. In the optimized configurations, the linear far-field reflectance spectra clearly show mode splitting near the excitonic frequency, which is an initial indicator for absorption due to possible SC. In nanohole array based structures, the interaction between the resonances is complicated, and as a result, plasmonic effects such as resonance red shifts (due to the presence of WS$_2$ in the hybrid structure), appearance of new resonances, multi-resonant coupling, spectral cut-offs etc. can be seen in addition to the spectral splitting itself. The simulated angular dispersion curves exhibit large mode splitting (Figure 4b,c,e,f) computed at the incident angle where the uncoupled resonances spectrally overlap. This splitting value is frequently associated to the Rabi splitting in plasmonic nanohole array systems[2,4], however the actual SC could be smaller due to the inherent plasmonic shift.

To show the superiority of the method, we also designed a periodic slits metasurface, whose geometry is easily tuned so that the spectral location of the far-field plasmonic resonance matches the excitonic one. In a slit array, we fulfill the momentum matching condition by choosing the array periodicity via the SPP dispersion formula (followed by a fine-tuning of the slit width and depth). The resulting structure is presented in Figure 5. This spectroscopically designed structure exhibits much smaller mode splitting compared with the optimized configurations, despite the superior Q factor of the plasmonic resonance. Note that the value of $E_m$ for the optimized configurations is higher than the one for the slits configuration which was not optimized via $E_m$ but rather using the spectroscopic approach.

To demonstrate the extensibility of our approach, we modify the $E_m$ function to support a specific functionality. We design a rectangular nanohole structure supporting both a strong coupling at $\lambda_{exc}$, and also a local field enhancement at another frequency (Figure 6). The modified optimization function is a weighted average of the mean field intensity at the two frequencies[28]. Such a design may be used in e.g. all-optical switching, or in 2-level perturbed strongly coupled systems[29].

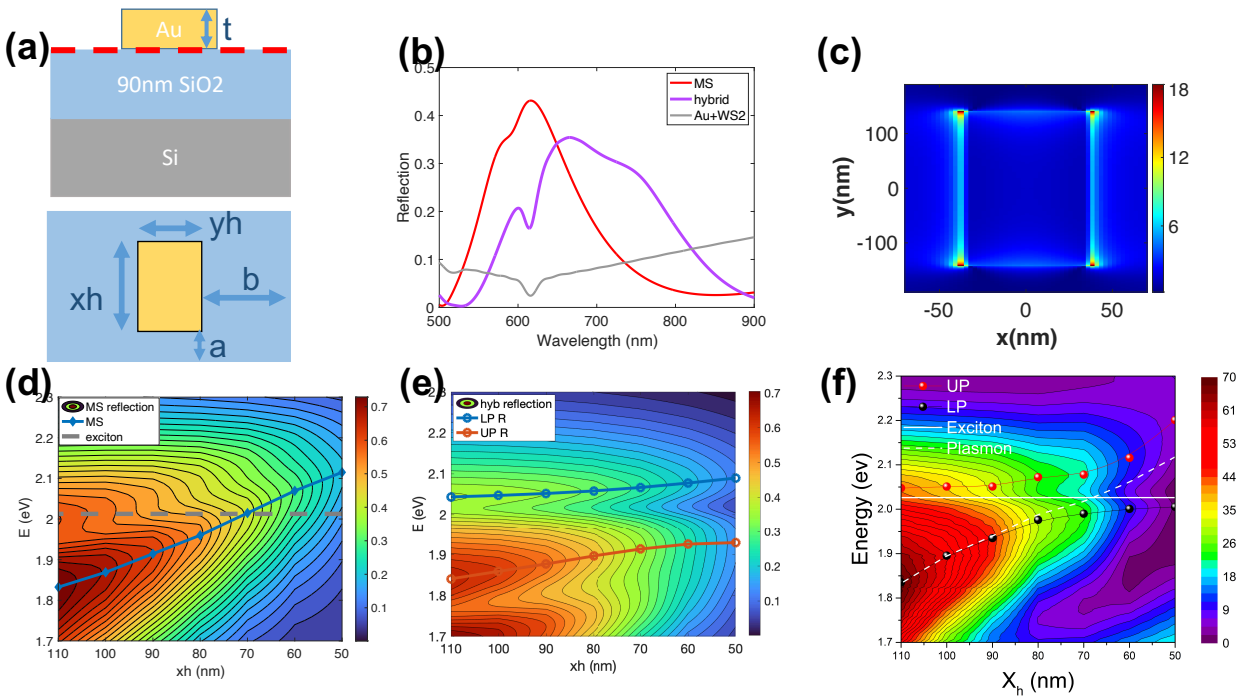

**Figure 2** Nanoantenna array optimization: **(a)** side and top view of a unit cell. The red dashed line shows the spatial location where the $E_m$ function is computed. Optimized configuration has $E_m$=1.7 and t=20nm, a=50nm, b=35nm, xh=70nm, yh=280nm. **(b)** Linear reflection spectra of the hybrid configuration and its uncoupled components. **(c)** The Ex field component of the optimized configuration. **(d,e)** Reflectance of the MS and the hybrid configuration with 0.5nm WS2, respectively, as a function of detuning parameter xh, showing the spectral peaks as colored dots. The Rabi splitting is 152 meV. (f) Experimental reflectance spectra of fabricated hybrid metasurfaces (see experimental methods).

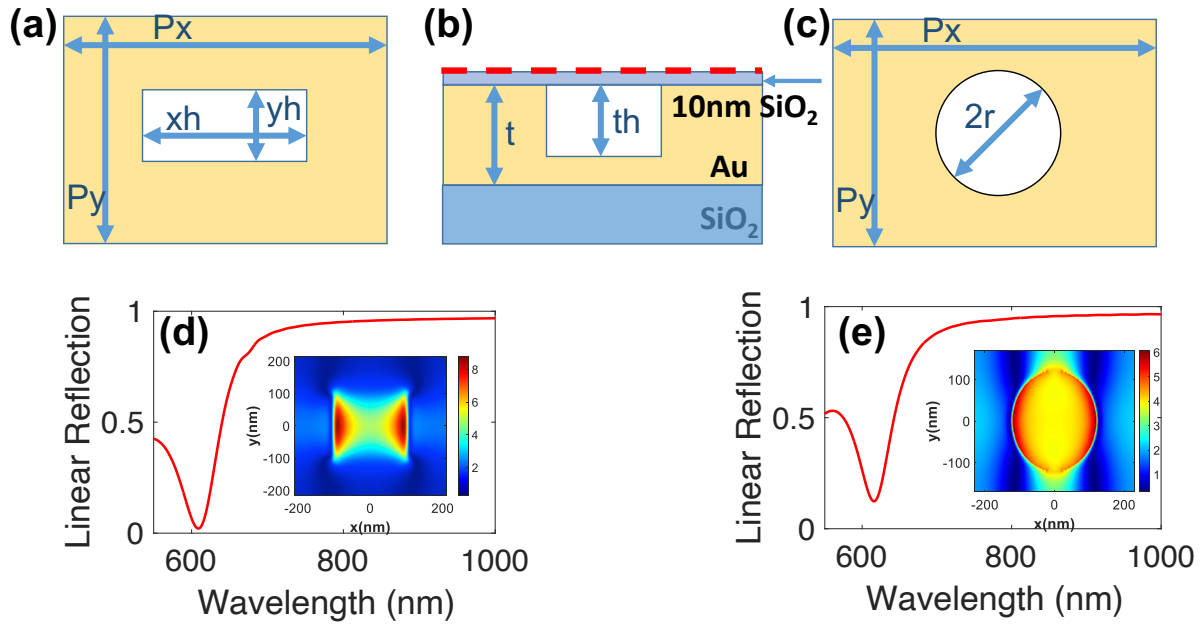

**Figure 3 (a-c)** Geometric parameter space for the nanocavity array MS (a single unit cell is shown): **(a)** Top view for rectangular geometry; **(b)** side view, where the red dashed line shows the spatial location where the $E_m$ function is computed; **(c)** top view for circular geometry. **(d,e)** Results of MS inverse design optimization for the rectangular and circular geometries, respectively. Incident light is linearly polarized in the XZ plane. For each configuration the reflectance spectra at normal incidence exhibits a resonance at or near $\lambda_{exc}$. Ex component in a unit cell right on top of the spacer layer[1] at normal incidence is shown inset. **(d)** optimal rectangular geometry ($E_m$=1.6), xh =yh =200nm, t=250nm, th=200nm, Px =420nm, Py =430nm. **(e)** optimal circular geometry ($E_m$=2.1), r=120nm, t=th=460nm, Px =460nm, Py =340nm;

---

[1] In nanohole arrays there is an additional 10nm of $SiO_2$ spacer. The spacer thickness is kept constant for fabrication convenience, but in fact it can be part of the optimization. The spacer provides an additional field enhancement and avoids quenching. On the other hand, plasmonic field enhancement is reduced as we move away from the MS.

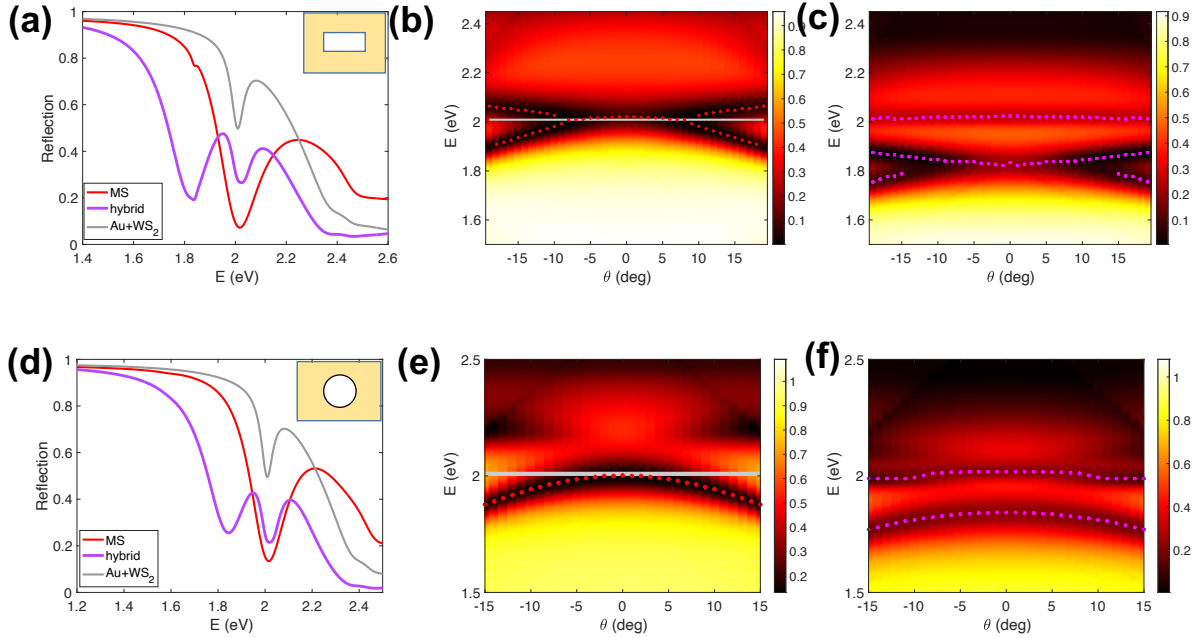

**Figure 4 (a-c)** Optimized rectangular geometry. **(a)** Reflectance spectra of the MS (red) and the exciton (grey) at normal incidence, both showing a resonance at $\lambda_{exc}$, together with the hybrid response (magenta). **(b)** Angle-resolved reflectance of the MS, showing the plasmonic resonance locations as red dots and the excitonic resonance as a grey solid line. $\theta$ is the angle of incidence. **(c)** Angle-resolved reflectance of the hybrid configuration, with LP/UP frequencies as magenta dots. Mode splitting is 177 meV, computed at $\theta = 7$ degrees. **(d-f)** Same as (a-c) for the optimized circular geometry. Mode splitting is 175 meV, computed at $\theta = 0$ degrees.

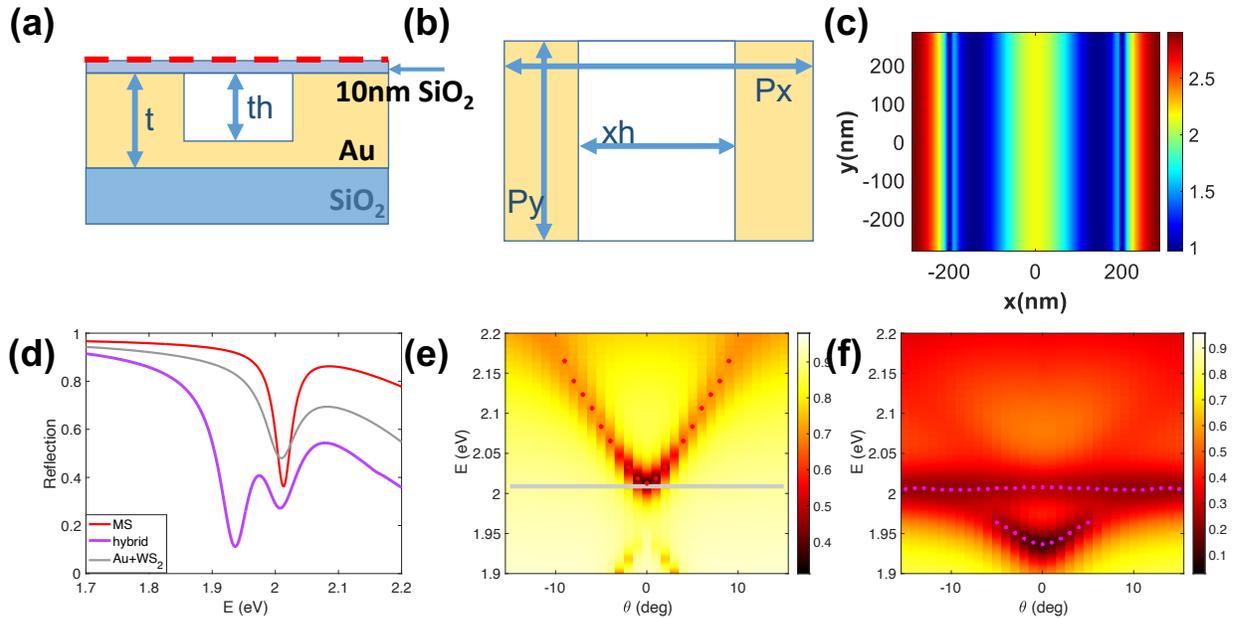

**Figure 5** The slit array configuration obtained by the spectroscopic approach. **(a)** Side and **(b)** top view of a unit cell. The parameters are xh=390nm, t=100nm. Px=Py=575nm, th=15nm ($E_m$=1.1). **(c)** Ex field component in a unit cell right on top of the spacer layer at normal incidence. **(d-f)** Same as (a-c) in **Figure 4**. The anticrossing point is $\theta = 0$. The mode splitting is 70 meV.

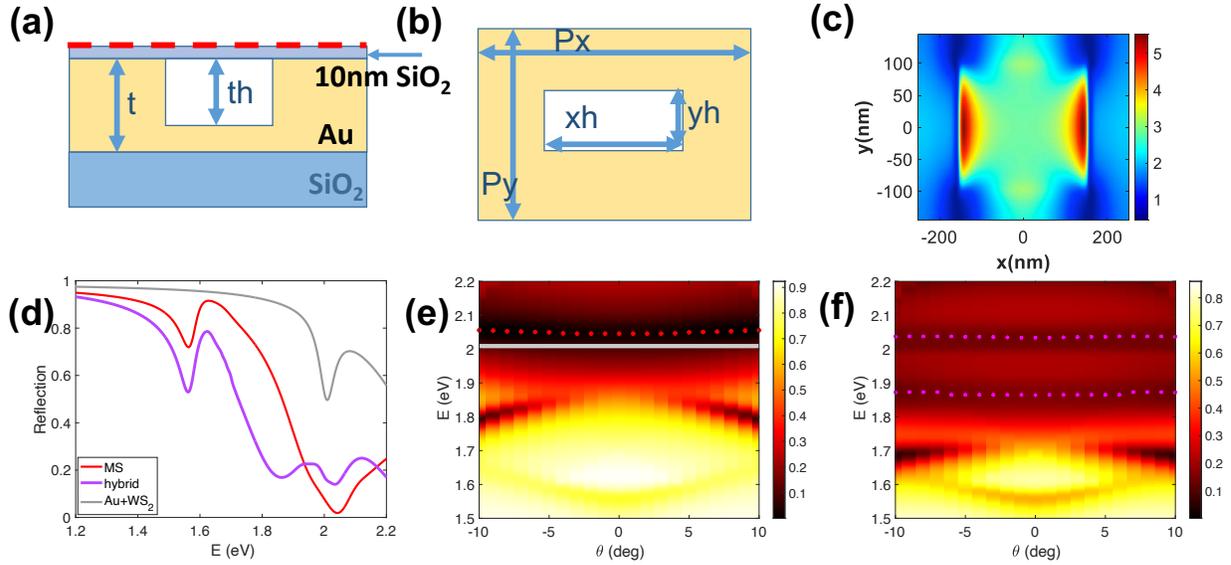

**Figure 6 Optimized multi-resonant rectangular geometry (a 2-plasmon-exciton system),** obtained by modifying the Em function to be a weighted average of the mean field at 612nm 800nm, with resp. weights 0.7 and 0.3 (modified Em=1.9). Optimal parameters are xh =300nm, yh =190nm, t=th=200nm, Px =500nm, Py=290nm. We obtain mode splitting of 168 meV, calculated after correction for nonzero detuning. The descriptions of the plots are the same as in **Figure 5**.

## Summary and discussion

In conclusion, we demonstrated an inverse design approach for obtaining large value of SC in nanometric systems. In particular, we showed that optimizing the mean electric field in the direction of the exciton dipole leads to a larger SC. For monolayer-thick $WS_2$ the optimized coupled nanoantenna array exhibit a very large Rabi splitting of 150 meV in simulation and 100 meV in measurement. For a 3nm-thick $WS_2$ we obtained mode splitting of 170 meV in the optimized nanohole array configurations, compared with less than 70 meV in the spectroscopically-designed slits array.

Our method can be applied in additional nanometric setups. For example, in metal-on-mirror or nanogap cavities[23] or stacked 2d materials (such as heterostructures[30]), one can use the same $E_m$ except using the out-of-plane field component in the definition. Another possible modification is to use quantum dot emitters instead of TMDC[31,32], which only requires a change in the integration boundaries for the computation of $E_m$. In yet another direction, one can decrease hot-spot size in order to minimize the number of coupled emitters, which, according to current research efforts, can lead to possible applications in quantum information processing[1]. Finally, our $E_m$ function can be easily used in alternative inverse design methods such as machine learning techniques which can drastically improve optimization speeds[33].

# Experimental methods

**Fabrication**

Gold nanoantenna arrays were fabricated by electron beam lithography on CVD-grown tungsten disulfide. A high-quality (99.9995% purity), full area coverage WS2 monolayer on a 90 nm SiO2/Si substrate (10x10 mm2) was purchased from 2D semiconductors. A PMMA 950k A2 resist film of thickness equal to 140 nm was spin-coated on the WS2 sample and pre-baked at 180 ℃. The patterns were exposed by a 50 kV electron beam with a current of 200 pA and dose of 600 µC/cm2 and developed on a (1:3) MIBK/IPA solution for 60 s. A 20-nm thick gold layer was deposited onto the sample by e-beam evaporation at a rate of 0.5 Å/s. The sample was left overnight in acetone at 45 ℃ for lift-off, cleaned in isopropanol and N2 blow-dried.

**Optical measurements**

The reflectance spectra of the arrays were measured on a microscope using a quartz-tungsten-halogen broadband light source. The light was linearly polarized along the horizontal axis by a calcite polarizer. The reflected light was collected by a fiber-coupled spectrometer (Ocean Optics USB 4000) with a resolution of 1.235 nm (FWHM). To obtain a reference spectrum, a high-reflectivity (R > 96%) silver mirror was placed in the sample stage.

# Supplementary information

## Optimizing with Genetic algorithm

In optimization, one usually searches for parameter values $x_0, y_0, z_0, \ldots$ at which some function $f(x, y, z, \ldots,)$ attains its extreme value. Many specialized methods exist depending on the analytic properties of the function, quickly converging to the relevant regions in the parameter space and saving enormous amounts of time (especially when the number of variables is large). Finding optimal configurations is a special optimization problem. In our case, the function to be maximized is related to the near field response obtained from running an FDTD simulation, and the parameters are related to hole/antenna shape, size, array periodicity and film thickness.

Genetic algorithms are inspired by the biological evolution[26]. The process is initialized by randomly selecting an initial population – the "first generation" – of "individuals" ("members"), where each individual represents a point in parameter space (the values of the parameters are its "genes"). The algorithm proceeds by evaluating each individual according to the specified objective function. Those individuals receiving the larger values (more "fit to survive") are then used to construct the next generation of individuals, by using various "biological operators":
1. "Mutation": the genes of a single individual are randomly modified (within a predefined range).
2. "Cross-breeding": the genes of some two individuals ("parents") are mixed to create a new individual ("offspring").

This creates a continuous improvement process, typically converging to an optimized individual (no more improvement when moving to the next generation), which is, hopefully, the best possible one inside the defined parameter space.

We chose to use genetic algorithm because it is efficient precisely in cases when we don't know the derivative (or it doesn't exist), or the function is not even known to be continuous. This is in contrast to other methods such as steepest descent.

## $WS_2$ thickness and implications for SC

It is well known that the number of TMDC atomic layers has significant effect on the coupling strength. For the nanoslit array design obtained by the spectroscopic approach, we have observed that at least 5 atomic layers of $WS_2$ was necessary to observe spectral splitting in the hybrid configuration (Supplementary Figure S1). Therefore, in order to compare this configuration with the 3 optimized nanocavity array designs, we have used 3nm-thick layer[34] of $WS_2$ during validation. This choice also plays well with both fabrication and simulation constraints. Indeed, in order to transfer a flake on top of the MS, one usually uses exfoliation, which is challenging for less than 5 atomic layers. On the other hand, simulating angle dispersion requires using Bloch boundary conditions (BC), and therefore reducing the $WS_2$ thickness even further would require even smaller mesh size and, as a result, highly demanding computational resources. In our calculations, we use the dielectric function of monolayer $WS_2$, which has a sharp resonant peak at the excitonic transition. Consequently, all our results below are approximately valid for 5 monolayers on top of each other. On the other hand, if a 3nm-thick bulk of $WS_2$ is transferred, we expect some deviations from our calculations in practice. Furthermore, since the plasmonic fields are not changing significantly in the 3nm range from the interface, we expect our optimization to produce optimal strongly coupled structures in this case as well.

In simulations of nanoantenna arrays, $WS_2$ thickness was taken to be 0.5nm, due to a different fabrication technique (e-beam lithography of the antenna on top of a commercially-bought large area CVD-grown monolayer $WS_2$, without the need for transfer). Even with the required increase in mesh accuracy, the dispersion simulations become significantly less demanding with periodic BC instead of Bloch BC.

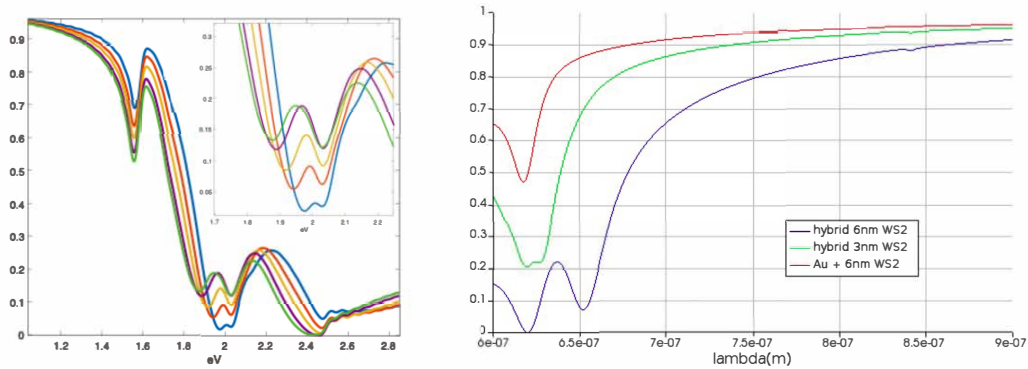

**Figure S1 Calculated dependence of the optical response of the hybrid structure, on the number of WS2 monolayers, (a) for the circular hole pattern (b) for nanoslits. As expected, the SC increases with the number of excitons participating in the interaction, but because of the decreasing overlap with the plasmonic mode (which is evanescent), the SC saturates.**

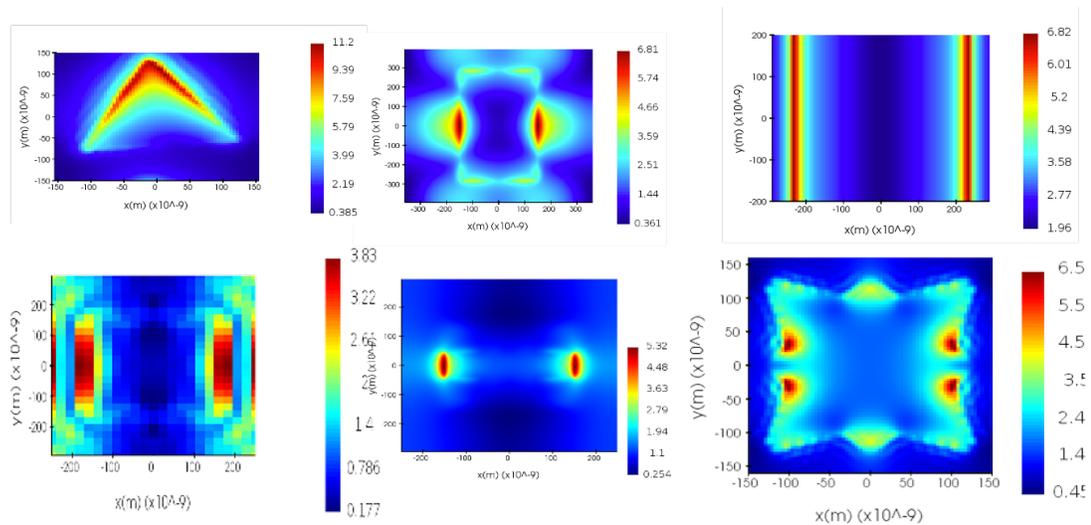

**Figure S2 Near field distribution of different plasmonic MS, giving the same far field response. Here we see few examples of near field distributions in unit cells corresponding to different structures - just the MS without the 2D material. We can see that beside the variation in the intensity of the fields, there is also a variety of field distributions.**